\documentstyle[12pt,aasms4]{article}
\begin{document}

\vspace*{0.5cm}

\title{Neutrino Pair Annihilation in the Gravitation of Gamma Ray Burst Sources}

\author{Katsuaki Asano and Takeshi Fukuyama}

\vspace{2cm}

\affil{Department of Physics, Ritsumeikan University\\
Kusatsu, Shiga 525-8577, Japan}

\affil{\footnotesize e-mail: sph10001@se.ritsumei.ac.jp}

\vspace{2cm}

\baselineskip=16pt

\abstract{We study semianalytically
the gravitational effects on neutrino pair annihilation
near the neutrinosphere and around the thin accretion disk.
For the disk case, we assume that the accretion disk is isothermal
and that the gravitational field is dominated by the Schwarzschild black hole.
General relativistic effects are studied only near the rotation axis.
The energy deposition rate is enhanced by
the effect of orbital bending toward the center.
However, the effects of the redshift and gravitational trapping
of the deposited energy reduce the effective energy of the gamma ray bursts' source.
Although each effect is substantial,
the effects partly cancel one another.
As a result, the gravitational effects do not substantially change the
energy deposition rate for either the spherical symmetric case or the disk case.}

\vspace{0.5cm}

\noindent{\it Subject headings}: accretion, accretion disks---black hole physics---gamma rays: bursts

\newpage

\section{INTRODUCTION}

\indent

The relativistic fireball (Shemi \& Piran 1990; Rees \& M\'esz\'aros 1992;
M\'esz\'aros \& Rees 1993; Sari \& Piran 1995; Sari, Narayan \& Piran 1996)
is  one of the most promising models of gamma ray bursts (GRBs).
However, even if the fraction of the baryon rest energy is only $10^{-3}$
in the fireball, the relativistic bulk flow, which is
indispensable to GRBs, cannot be realized.
Notwithstanding the very high energy phenomenon ($10^{52}$ ergs), the baryon
density in the fireball must be extremely small.
This is the famous baryon contamination problem and still remains unsolved.
Thus the central engine of GRBs is still beyond
deep mist.
The source of GRBs may be one super massive
(failed) supernovae  (Woosley 1993; Paczy\'nski 1998)
or may be a merger of two neutron stars or of a neutron star and a black hole
(e.g. Eichler et al. 1989; Narayan, Paczy\'nski \& Piran 1992;
M\'esz\'aros \& Rees 1992a; Katz 1997; Ruffert \& Janka 1998, 1999).

In these compact high energy objects,
the neutrino-antineutrino annihilation
into electrons and positrons
(hereafter neutrino pair annihilation)
is a possible and important candidate to explain the energy source of GRBs
(Paczy\'nski 1990; M\'esz\'aros \& Rees 1992b; Janka \& Ruffert 1996;
Ruffert et al. 1997; Ruffert \& Janka 1998, 1999).
Motivated by the delayed explosion of Type II supernovae,
the energy deposition rate due
to the neutrino pair annihilation above the neutrinosphere
has been calculated
(Goodman, Dar \& Nussinov 1987; Cooperstein, Van Den Horn \& Baron 1987;
Berezinsky \& Prilutsky 1987).
The energy deposition rate is proportional to $r^{-8}$ ($r$
is the distance from the center of the neutrinosphere) for a large $r$,
and almost all deposition occurs near the neutrinosphere.
As they themselves noted in their paper,
Goodman et al. (1987) neglected the general
relativistic effects on the energy deposition rate, which may change
their numerical value seriously.
In simulations of the neutrino pair annihilation rate,
it is very important to confirm whether or not the energy deposition
rate is altered or not by the gravitational effects.
In the recent study, Salmonson \& Wilson (1999)
concluded that the energy deposition rate in Type II
supernovae is enhanced about 4 times as a result of the gravitational effects.
We must check whether or not their results can be applied to the central engine
of GRBs.

One of the most probable candidates for the central engine of GRBs
is the accretion disk around a black hole (Woosley 1993; Popham, Woosley \& Fryer 1999;
MacFadyen \& Woosley 1999; Ruffert \& Janka 1999).
The system of an accretion disk and a black hole may be formed
by the merging of two neutron stars,
the merging of a black hole and a neutron star,
or the failed supernovae.
In general, the baryon density has the lowest value along the rotation axis
just above the black hole (e.g. see Ruffert \& Janka 1999).
This region might be a key to resolving the baryon contamination problem.
The hot accretion disk emits neutrinos and antineutrinos.
The energy deposited in the lowest density region is a candidate for
the central engine of GRBs.
Using hydrodynamic simulations,
Ruffert \& Janka (1999) showed that the neutrino pair annihilation deposits 
energy in the vicinity of the torus at a rate of $(3-5)\times 10^{50}$ ergs ${\rm s^{-1}}$.
They concluded that the gravitational effect on the energy deposition rate
around the accretion disk is small.
We must supplement their results from the analytical side.

In various arguments on the energy deposition in the central engine of GRBs,
the order estimation of the deposited energy is sufficient, at least at present.
In this article, based on simple models, we study semianalytically
the gravitational effects on the energy deposition rate for two cases.
In one case neutrinos are emitted spherically symmetrically.
In the other case the hot accretion disk emits neutrinos.
We have derived the gravitational effects on the former case independently of
Salmonson \& Wilson (1999).
Some differences of our work from Salmonson \& Wilson in the formulation,
the interpretation of the energy deposition, and the additional factor are
mentioned.
As for the disk case, we assume that the accretion disk is isothermal
and that the gravitational field is dominated solely by the central
Schwarzschild black hole.
These assumptions enable us to treat the energy deposition around the disk
semianalytically.
Thus, in both two cases gravitation is described by the Schwarzschild metric,
and the essential differences between the two cases
come from the shape of the neutrino emitters.

The gravitational effects consist of three factors:
they are the bending of neutrino
trajectories, the gravitational redshift, and the trapping of deposited
energy into the central gravitational source.
We show that the energy deposition rate is indeed enhanced rather crucially by
the effect of neutrino bending.
However, it is also shown that the gravitational redshift and the trapping
of the deposited energy reduce this enhancement.
As a result, the gravitational effects do not substantially change the
energy deposition rate for either the spherical symmetric case or the disk case.

This paper is organized as follows. In section two we investigate
neutrino pair annihilation near the neutrinosphere.
The same process around the accretion disk is
discussed in section three.  The last section is devoted to conclusions.

\section{NEUTRINO PAIR ANNIHILATION NEAR THE NEUTRINOSPHERE}

\indent

In this section we study the general relativistic effects on
neutrino pair annihilation near the neutrinosphere.
This study has been already done by Salmonson \& Wilson (1999).
Using another method, we formulate the same problem independently of
the work of Salmonson \& Wilson.
Some alterations in the interpretation of the energy deposition
in Salmonson \& Wilson are mentioned.

The number of reaction, $\nu+\bar{\nu} \to e^+ +e^-$,
per unit volume per unit time (Goodman, Dar \& Nussinov 1987) is written as
\begin{equation}
\frac{dN(\mbox{\boldmath$r$})}{dt dV}=
\int \int f_{\nu}(\mbox{\boldmath$p$}_{\nu}, \mbox{\boldmath$r$})
f_{\bar{\nu}}(\mbox{\boldmath$p$}_{\bar{\nu}}, \mbox{\boldmath$r$})
\sigma \left| \mbox{\boldmath$v$}_{\nu}-\mbox{\boldmath$v$}_{\bar{\nu}} \right|
d^3 \mbox{\boldmath$p$}_{\nu} d^3 \mbox{\boldmath$p$}_{\bar{\nu}}.
\label{first}
\end{equation}
Here $f_{\nu}$ ($f_{\bar{\nu}}$) is the number density of neutrinos
(antineutrinos) in phase space,
$\mbox{\boldmath$v$}_{\nu}$ ($\mbox{\boldmath$v$}_{\bar{\nu}}$)
is the velocity of neutrinos (antineutrinos),
and $\sigma$ is the rest-frame cross section.
The left handside of equation (\ref{first}) is Lorentz invariant,
since both the numerator, $dN$, and denominator, $dt dV=\sqrt{-g} d^4 x$,
are Lorentz invariant.
Since $f_{\nu}$ and $d^3 \mbox{\boldmath$p$}_{\nu}/\varepsilon_{\nu}$
(where $\varepsilon_{\nu}$ is the proper energy of neutrinos) of the right handside
are also Lorentz invariant,
$\varepsilon_{\nu} \varepsilon_{\bar{\nu}}
| \mbox{\boldmath$v$}_{\nu}-\mbox{\boldmath$v$}_{\bar{\nu}} | \sigma$
should be Lorentz invariant.
The latter is written in a manifest Lorentz-invariant form as
$ \sigma c^3 (p_{\nu} \cdot p_{\bar{\nu}})$,
where $(p_{\nu} \cdot p_{\bar{\nu}})$ is the inner product of the 4-momenta.
The standard model predicts that the cross section is expressed as
\begin{equation}
\sigma=2 c^2 K G_{\rm F}^2 (p_{\nu} \cdot p_{\bar{\nu}}),
\end{equation}
where the dimensionless parameter $K$ is written as
\begin{eqnarray}
K(\nu_\mu \bar{\nu_\mu})&=&K(\nu_\tau \bar{\nu_\tau})=
\frac{1-4 \sin^2{\theta_{\rm W}}+8\sin^4{\theta_{\rm W}}}{6 \pi}, \nonumber \\
K(\nu_{\rm e} \bar{\nu_{\rm e}})&=&
\frac{1+4 \sin^2{\theta_{\rm W}}+8\sin^4{\theta_{\rm W}}}{6 \pi}.
\end{eqnarray}
Here the Fermi constant $G_{\rm F}^2=5.29 \times 10^{-44} {\rm cm^2\, MeV^{-2}}$ and
the Weinberg angle $\sin^2{\theta_{\rm W}}=0.23$.

Let us incorporate the effects of gravitational force due to the neutron star
or black hole on the neutrino pair annihilation rate.
We assume that the gravitational field is described by the Schwarzschild
metric:
\begin{equation}
ds^2=g_{i j} dx^i dx^j=
\left( 1-\frac{r_{g}}{r} \right) c^2 dt^2-\frac{1}{1-\frac{r_{g}}{r}} dr^2
-r^2 \left( d\theta^2-\sin^2{\theta} d\varphi^2 \right),
\end{equation}
where $r_{g}=2GM/c^2$ is the Schwarzschild radius.
In this field the eikonal for a massless particle 
(Landau \& Lifshitz 1979) is written as
\begin{equation}
\psi=-\omega_0 t+L \varphi+\psi_r(r),
\label{eikonal}
\end{equation}
where $\omega_0$ and $L$ are constants.
$\psi_r(r)$ satisfies the equation
\begin{equation}
\frac{\partial \psi_r(r)}{\partial r}=
\sqrt{\frac{\omega_0^2}{c^2} \left( 1-\frac{r_{g}}{r} \right)^{-2}-\frac{L^2}{r^2}
\frac{1}{1-\frac{r_{g}}{r}}}.
\end{equation}
From equation (\ref{eikonal}),
we can obtain the momentum of a neutrino by
$p_{i}=\hbar \frac{\partial \psi}{\partial x^i}$.

Let us consider a neutrino and an antineutrino moving on the same surface,
$\theta=\pi/2$.
In this case, the inner product of the two particles is written by
\begin{eqnarray}
(p_{\nu} \cdot p_{\bar{\nu}})&=&g^{ij} p_{\nu i} p_{\bar{\nu} j} \\
&=& \frac{\varepsilon_{\nu} \varepsilon_{\bar{\nu}}}{c^2}
\left( 1-\sqrt{1-\left( \frac{\rho_{\nu}}{r}
\right)^2 \left( 1-\frac{r_{g}}{r} \right)}
\sqrt{1-\left( \frac{\rho_{\bar{\nu}}}{r} \right)^2
\left( 1-\frac{r_{g}}{r} \right)} \right. \nonumber \\
&& \left. -\frac{\rho_{\nu} \rho_{\bar{\nu}}}{r^2}
\left(1-\frac{r_{g}}{r} \right) \right),
\label{product}
\end{eqnarray}
where
\begin{equation}
\rho_{\nu} \equiv \frac{c L_\nu}{\omega_{0 \nu}}.
\end{equation}
The proper energy of the neutrino has been written as
\begin{equation}
\varepsilon_{\nu} = \frac{\hbar \omega_{0 \nu}}{\sqrt{1-\frac{r_{g}}{r}}}
\equiv \frac{\varepsilon_{0 \nu}}{\sqrt{1-\frac{r_{g}}{r}}},
\end{equation}
where $\varepsilon_{0 \nu}$ is the energy observed at infinity.
Thus the proper energy is redshifted, as is well known.
If we define an angle $\theta_{\nu}$ as
\begin{equation}
\sin{\theta_{\nu}} = \frac{\rho_{\nu}}{r} \sqrt{1-\frac{r_{g}}{r}},
\label{sin}
\end{equation}
equation (\ref{product}) becomes a simple and natural form,
\begin{equation}
(p_{\nu} \cdot p_{\bar{\nu}})=\frac{\varepsilon_{\nu} \varepsilon_{\bar{\nu}}}{c^2} 
\left( 1-\cos{(\theta{\nu}-\theta{\bar{\nu}})} \right).
\end{equation}
The angle $\theta_{\nu}$ ($\theta_{\bar{\nu}}$)
represents the angle between $\mbox{\boldmath$p$}_{\nu}$ ($\mbox{\boldmath$p$}_{\bar{\nu}}$)
and the position vector $\mbox{\boldmath$r$}$ (see Figure 1).
We assume that the neutrinosphere emits neutrinos and antineutrinos isotropically.
Then we can write the number densities as
$f_{\nu}(\mbox{\boldmath$p$}_{\nu}, \mbox{\boldmath$r$})
d^3 \mbox{\boldmath$p$}_{\nu}=n(\varepsilon_{\nu})
\varepsilon_{\nu}^2 d \varepsilon_{\nu} d \Omega$.
Because $\rho_{\nu}$ is constant along a neutrino ray,
the maximum angle, $\theta_{\rm M}$, is obtained by
substituting $\pi/2$ for $\theta_{\nu}$ at the radius of the
neutrinosphere, $R_{\nu}$, in equation (\ref{sin}).
Thus we obtain
\begin{equation}
\sin{\theta_{\rm M}}=\frac{R_{\nu}}{r} \sqrt{\frac{1-\frac{r_{g}}{r}}
{1-\frac{r_{g}}{R_{\nu}}}}.
\label{sM}
\end{equation}
The effect of the orbital bending is apparent in this equation.
Until now we have discussed the maximum angle on the surface of $\theta=\pi/2$.
In general cases, the angles between $\mbox{\boldmath$p$}_{\nu}$
and $\mbox{\boldmath$r$}$ or the inner product, $(p_{\nu} \cdot p_{\bar{\nu}})$,
are expressed by the two angles, $\theta_{\nu}$
and $\varphi_{\nu}$.
From the symmetry, the behaviour of $\theta_{\rm M}$ is obviously
the same as described in equation (\ref{sM}),
and $\varphi_{\nu}$ varies from $0$ to $2 \pi$.

Using the effective temperature of the neutrinosphere,
$T_{\rm eff}=T_0/\sqrt{g_{00}}$ with a constant $T_0$,
we can write the density
\begin{equation}
n(\varepsilon_{\nu})=\frac{g_{\nu}}{(hc)^3} \frac{1}{\exp{\left( 
\frac{\varepsilon_{\nu}}{k T_{\rm eff}}\right)}+1},
\end{equation}
where $g_{\nu}$ is a statistical factor
($g_{\nu}=1$ for a neutrino).
$\varepsilon_{\nu}/(k T_{\rm eff})$ is constant along a neutrino ray,
since the redshift is cancelled out.
Thus $n(\varepsilon_{\nu})$ is conserved along a neutrino ray
in accordance with Liouville's theorem in curved spacetime
(Misner, Thorne \& Wheeler 1975).
From the above formulation, one can find
\begin{equation}
\frac{dN(\mbox{\boldmath$r$})}{dt dV}=
2 c K G_{\rm F}^2 F(r) \int \int d \varepsilon_{\nu} d \varepsilon_{\bar{\nu}}
n(\varepsilon_{\nu}) n(\varepsilon_{\bar{\nu}})
\varepsilon_{\nu}^3 \varepsilon_{\bar{\nu}}^3,
\label{num}
\end{equation}
where the dimensionless factor $F(r)$ is written by
\begin{eqnarray}
F(r)&=&\int_0^{\theta_{\rm M}} d \theta_{\nu} \sin{\theta_{\nu}}
\int_0^{\theta_{\rm M}} d \theta_{\bar{\nu}} \sin{\theta_{\bar{\nu}}}
\int_0^{2 \pi} d \varphi_{\nu} \int_0^{2 \pi} d \varphi_{\bar{\nu}} \nonumber \\
&& \times \left( 1-\sin{\theta_{\nu}} \sin{\theta_{\bar{\nu}}}
\cos{(\varphi_{\nu}-\varphi_{\bar{\nu}})}-\cos{\theta_{\nu}} \cos{\theta_{\bar{\nu}}}
\right)^2 \label{Fr} \\
&=& \frac{2 \pi^2 (1-X)^4}{3} (X^2+4X+5),
\end{eqnarray}
where
\begin{equation}
X = \sqrt{1-\left( \frac{R_{\nu}}{r} \right)^2
\frac{1-\frac{r_{g}}{r}}{1-\frac{r_{g}}{R_{\nu}}}}.
\end{equation}
In our assumption, the energy deposited by the neutrino pair annihilation
is propagated outward as a fireball or a shock wave,
and observed as a GRB by a distant observer.
Thus the energy we need to calculate is $\varepsilon_{0 \nu}$,
not the proper energy $\varepsilon_{\nu}$.
In this case the energy deposition rate is obtained by putting a factor
$(\varepsilon_{0 \nu}+\varepsilon_{0 \bar{\nu}})$ in the integrand
in equation (\ref{num});
\begin{eqnarray}
\frac{dE_0 (\mbox{\boldmath$r$})}{dt dV}&=&
\frac{2 c K G_{\rm F}^2}{\left(  1-\frac{r_{g}}{r} \right)^4}
\int_0^{\infty} \int_0^{\infty} d \varepsilon_{0 \nu} d \varepsilon_{0 \bar{\nu}}
\nonumber \label{depo} \\
&& \times n(\varepsilon_{0 \nu}) n(\varepsilon_{0 \bar{\nu}})
\varepsilon_{0 \nu}^3 \varepsilon_{0 \bar{\nu}}^3
(\varepsilon_{0 \nu}+\varepsilon_{0 \bar{\nu}}) F(r) \\
&=& \frac{21 \pi^4}{4} \zeta(5) \frac{K G_{\rm F}^2 g_{\nu}^2}{h^6 c^5}
\frac{\left( 1-\frac{r_{g}}{R_{\nu}} \right)^{\frac{9}{2}}}
{\left(  1-\frac{r_{g}}{r} \right)^4} (k T_{\rm eff})^9 F(r).
\label{depo2}
\end{eqnarray}
The integrals for $\varepsilon_{0 \nu}$ and $\varepsilon_{0 \bar{\nu}}$
should be defined in the range in which the total energy
produced by the pair annihilation is larger than the mass of created electrons,
and smaller than the masses of weak bosons.
Here we have approximated the integrals as expressed in equation (\ref{depo})
in the same manner as Salmonson \& Wilson (1999) did.
This is because the cross section decreases with the energy of neutrinos,
and the number of neutrinos whose energy is larger than the masses of weak bosons
is also very small in our assumption ($k T_{\rm eff}$ is of the order of several MeV).
The factor $( 1-r_{g}/R_{\nu})^{9/2} /(  1-r_{g}/r )^4$ represents
the effect of the gravitational redshift,
and $F(r)$ includes the effect of the orbital bending.

As is understood from the Lorentz invariant, $dtdV=\sqrt{-g} d^4 x$,
if we integrate equation (\ref{depo2}) over proper volume,
$dV'=\sqrt{-g_{rr} g_{\theta \theta} g_{\varphi \varphi}} dr d\theta d\varphi$,
we can obtain the total energy deposition per unit proper time,
$d \tau=\sqrt{g_{00}} dt$.
It is natural to evaluate the energy deposition rate
by the world time $dt$ for a distant observer.
We integrate over the volume,
$dV=\sqrt{g_{\theta \theta} g_{\varphi \varphi}} dr d\theta d\varphi$.
Thus the energy deposition per unit world time is expressed as
\begin{equation}
\frac{dE_0}{dt}=
\frac{21 \pi^4}{4} \zeta(5) \frac{K G_{\rm F}^2 g_{\nu}^2}{h^6 c^5} (k T_{\rm eff})^9
\left( 1-\frac{r_{g}}{R_{\nu}} \right)^{\frac{9}{2}}
\int_{R_{\nu}}^{\infty} dr 4 \pi r^2
\frac{F(r)}{\left(  1-\frac{r_{g}}{r} \right)^4} C(r),
\label{result}
\end{equation}
where we have put a factor,
\begin{equation}
C(r)=\frac{1}{2} \left( 1+\sqrt{1-\frac{27}{4} \left( \frac{r_{g}}{r} \right)^2
\left( 1-\frac{r_{g}}{r} \right) } \right),
\end{equation}
in the integrand.
This is the escape probability of the deposited energy at $r$
from the gravitational attraction
(Chandrasekhar 1983; Shapiro \& Teukolsky 1983; Ruffert \& Janka 1999).
The electrons, positrons and photons which are captured by the gravitational attraction
cannot contribute to the energy source of GRBs.
Apart from $C(r)$,
the radial profile in the integrand of equation (\ref{result}) 
is different from those of Salmonson \& Wilson (1999),
since Salmonson \& Wilson calculated the proper energy deposition per
unit proper time.
Of course, the results of Salmonson \& Wilson are not mistakes for the estimate of
the energy deposition rate in supernovae.
For the source of GRBs, however,
equation (\ref{result}) is adequate.

Let us investigate the effects of the redshift, orbital bending and gravitational capture.
We integrate equation (\ref{result}) and obtain
the energy deposition rate for $\nu_{e}$ as
\begin{equation}
\frac{dE_0}{dt}=1.27 \times 10^{42} \left( \frac{k T_{\rm eff}}{1 {\rm MeV}} \right)^9
\left( \frac{R_{\nu}}{10 {\rm km}} \right)^3 f \quad \mbox{ergs ${\rm s^{-1}}$},
\label{Gf}
\end{equation}
where the dimensionless factor $f$ expresses the effects of the general relativity
($f=1$ when we neglect the gravitation).
The energy deposition rates for $\nu_{\mu}$ and $\nu_{\tau}$ are 0.64 times
equation (\ref{Gf}).
We numerically estimate $f$ including the effects of the redshift only,
the orbital bending only, or both redshift and orbital bending.
Last, the total effects of the redshift, orbital bending, and gravitational capture
are calculated.
The results are listed in Table 1.
As Table 1 or equation (\ref{result}) indicates,
the effect of the redshift reduces the energy deposition rate,
and the effect of the orbital bending increases it.
Although each effect, that of the redshift and that of orbital bending, is substantial,
the effects partly cancel each other.
As a result, the order of the energy deposition rate for
the most probable case, $R_{\nu}/r_g=2.5$, is not altered.
When we neglect the general relativistic effects, the energy deposition rate
increases by 1.3 times as $R_{\nu}$ becomes 10\%
larger, and  also increases by 2.4 times as the temperature becomes 10\%
higher.
Therefore, the gravitational effects are not so large in comparison with
the errors due to the uncertainties of $R_{\nu}$ or $T_{\rm eff}$,
and are overwhelmed by them.
The effect of the gravitational capture becomes important
as $R_{\nu}/r_g$ decreases.
As is plotted in Figure 2, in the cases of both the presence and absence
of gravitation, the energy deposition mainly occurs near the neutrinosphere.

Salmonson \& Wilson (1999) concluded that the effects of gravity
enhance the energy deposition rate up to a factor of more than
4 for $R_{\nu} \le 2.5 r_g$.  However, our results
show that the gravitational effects reduce the energy deposition rate.
This discrepancy survives even if we omit the escape factor $C(r)$.
The proper energy deposition per unit proper time is enhanced
by both the effects of the redshift and that of orbital bending.
Additionally, Salmonson \& Wilson expressed the general relativistic
effects with the fixed neutrino luminosity at infinity $L_{\infty}$,
whereas we have done so with the local physical quantity $T_{\rm eff}$.
Therefore, an additional factor coming from the redshift of the luminosity
($L(R_{\nu}) \propto T_{\rm eff}^4$, $L_{\infty}=(1-{r_g/R_{\nu}})L(R_{\nu})$)
enhances the energy deposition rate in the work of Salmonson \& Wilson.
However, the quantity $L_{\infty}$ of GRBs is not directly observable at present.
It is more natural to study the effects for the given local parameters,
$T_{\rm eff}$ or $L(R_{\nu})$,
which is restricted or provided by models of the central engine.

\section{NEUTRINO PAIR ANNIHILATION AROUND THE ACCRETION DISK}

\indent

In this section we investigate the energy deposition rate around the accretion disk.
In order to simplify our formulation,
we assume that the accretion disk is isothermal
and that the gravitational field is dominated
by the central Schwarzschild black hole.
We neglect the rotation of the black hole.
The accretion disk is assumed to be thin, and its self-gravitational effects
are neglected.
Of course, these idealizations may be far from the case of the realistic accretion disk.
However, we consider that this simple method is sufficient for qualitatively studying
the gravitational effects on the energy deposition rate.
In this case the equation of the energy deposition rate is
the same as equation (\ref{depo2}) provided that $F(r)$
is replaced by $F(r,\theta)$ (it will be given below).
The effect of the gravitational redshift can be easily incorporated,
whereas the formulation of the neutrino bending is difficult to do
because the accretion disk emits neutrinos anisotropically.

First, we calculate the dimensionless factor $F(r, \theta)$
without the effect of gravity.
The accretion disk is placed on the equatorial plane, $\theta=\pi/2$.
The black hole is at the origin, and we consider
a point $P=(r,\theta,0)$ where pair annihilations occur (see Figure 3).
A neutrino is emitted from an arbitrary point on the disk $S=(R,\pi/2,\varphi)$,
where $R$ is limited in the range from $R_{\rm in}$ to $R_{\rm out}$.
The neutrino emitted from $S$ travels straight and arrives at the point $P$.
Let us denote the angle components of the vector joining $S$ and $P$
by $(\theta_{\nu}, \varphi_{\nu})$.
They are given by
\begin{equation}
\cos{\theta_{\nu}}=\frac{r \cos{\theta}}{\sqrt{r^2+R^2-2 r R \sin{\theta} \cos{\varphi}}},
\end{equation}
\begin{equation}
\sin{\varphi_{\nu}}=\frac{-R \sin{\varphi}}
{\sqrt{r^2 \sin^2{\theta}+R^2-2 r R \sin{\theta} \cos{\varphi}}}.
\end{equation}
Thus $\theta_{\nu}$ and $\varphi_{\nu}$ are functions of $R$ and $\varphi$
for fixed $r$ and $\theta$.
The Jacobian $J \equiv \partial (\theta_{\nu}, \varphi_{\nu})/\partial (R, \varphi)$ is
\begin{equation}
J=\frac{r R \cos{\theta}}
{\sqrt{r^2 \sin^2{\theta}+R^2-2 r R \sin{\theta} \cos{\varphi}}
( r^2+R^2-2 r R \sin{\theta} \cos{\varphi})}.
\end{equation}
Consequently, we obtain $F(r,\theta)$ as
\begin{eqnarray}
F(r,\theta)&=&\int_{R_{\rm in}}^{R_{\rm out}} dR \int_{R_{\rm in}}^{R_{\rm out}} dR'
\int_0^{2 \pi} d \varphi \int_0^{2 \pi} d \varphi' J J' \nonumber \\
& \times & \sin{\theta_{\nu}} \sin{\theta_{\bar{\nu}}}
\left( 1-\sin{\theta_{\nu}} \sin{\theta_{\bar{\nu}}}
\cos{(\varphi_{\nu}-\varphi_{\bar{\nu}})}-\cos{\theta_{\nu}} \cos{\theta_{\bar{\nu}}}
\right)^2.
\label{Frtheta}
\end{eqnarray}

In equation (\ref{Frtheta}) we adopt $R_{\rm in}=3r_g$, the innermost stable orbit, and
$R_{\rm out}=10r_g$ as Woosely (1993) assumed. $F(r,\theta)$ derived from
the numerical integral of equation (\ref{Frtheta}) is plotted in Figure 4(a) and (b).
As is shown in these figures, the energy deposition rate is maximized in the
vicinity of the accretion disk, where $F(r,\theta) \simeq 30-33$.
The simulation of a neutron star merger by Ruffert \& Janka
(1999) showed that the Paczy\'nski-Wiita potential (Paczy\'nski \& Wiita 1980),
which mimics the effects of the general relativity,
gives a relatively more transparent disk for neutrinos than
that given by the Newtonian potential.
The profile of the energy deposition rate in the Paczy\'nski-Wiita potential
is similar to our analytical one depicted in Figure 4(a), which shows that the rate takes
its maximum value on the surface of the disk.
On the other hand, the simulated deposition rate in the Newtonian potential
is maximized near the rotation axis.
Let us calculate the energy deposition rate near the rotation axis,
where is the lowest baryon density region.
Thus we calculate in the region $\theta \leq \pi/4$
and obtain
\begin{equation}
\frac{dE_0}{dt}=5.22 \times 10^{43} \left( \frac{k T_{\rm eff}}{1 {\rm MeV}} \right)^9
\left( \frac{r_g}{10 {\rm km}} \right)^3 G f \quad \mbox{ergs ${\rm s^{-1}}$},
\label{Gf2}
\end{equation}
where the dimensionless quantity $G$ shows the relative contributions
from various regions in the absence of gravitation.
$G$ is normalized to unity when
we integrate over the volume for $\theta \leq \pi /4$ and $r=2r_g-10r_g$.
The values of $G$ in the other regions are summarized in Table 2, from
which we can obtain the energy deposition rate in the respective region.
We neglect the energy deposited inside $r=2 r_g$,
since the baryon density in this region is very high and
the energy contribution is small for the small volume and deposition rate.
In the case of the spherical emitter in section 2, the deposition rate decreases as
$r^{-8}$.
On the other hand, around the accretion disk, as is seen from Table 2, there remains a
marginal deposition rate even at regions relatively distant from the center.
Of course the deposition rate per unit volume at distant positions is small.
However, large volume results in a non-negligible contribution at the regions
distant from the center.

Untill now we have neglected the gravitational effects.
It is easy to incorporate the effects of the redshift and
trapping by the central gravitational source in the preceding arguments of this
section. However, the bending effect is difficult to treat, unlike the case
of the neutrinosphere, since the accretion disk emits neutrinos anisotropically.
Thus we are forced to make some approximation.
As is shown in Figure 4(b), the $\theta$ dependence of $F(r,\theta)$ is weak for 
small $\theta$.
We may set $F(r,\theta) \simeq F(r,0)$ for $\theta \leq \pi/4$.
In the absence of gravitation if we adopt this
approximation in the region $\theta \leq \pi /4$ and $r=2r_g-10r_g$,
we obtain $G=0.81$.
The exact value of $G$ is unity, and this approximation is not
necessarily satisfactory. However, this approximation may be sufficient for the order
estimate of the gravitational effects.

We can obtain $F(r,0)$ including the effect of orbital bending
with comparative ease,
since the geometry of this case maintain the symmetry.
A neutrino is emitted from the disk at $R$ and $\theta=\pi/2$,
and it arrives at a point at $r$ and $\theta=0$.
The nearest distance, $r_0$,
from the origin to the orbit of neutrinos (Landau \& Lifshitz 1979)
is numerically obtained from
\begin{equation}
\pi/2=\int_{\rm C} \frac{dr'}{r' \sqrt{\left( \frac{r'}{r_0} \right)^2
\left(1-\frac{r_g}{r_0} \right)-\left(1-\frac{r_g}{r'} \right)}}.
\end{equation}
Here, in the case in which a neutrino  passes through $r_0$
until it arrives at a point at $\theta=0$,
the integration for $r'$ is performed from $r_0$ to $R$ and $r$.
When the distance from the origin to the neutrino varies monotonically,
the integration is performed from the smaller
to the larger of $r$ and $R$.
We can get $\theta_{\nu}$ at $\theta=0$ numerically from $r_0$ and the following equation;
\begin{equation}
\sin{\theta_{\nu}}=\frac{r_0}{r} \sqrt{\frac{1-\frac{r_g}{r}}{1-\frac{r_g}{r_0}}}.
\end{equation}
The constant, $r_0$, or $\theta_{\nu}$, is a function of $r$ and $R$ in this case.
As is easily understood, a neutrino coming from $R_{\rm in}$ forms $\theta_{\rm m}$,
the minimum value of $\theta_{\nu}$, at $\theta=0$ and that from
$R_{\rm out}$ forms the maximum value of $\theta$, $\theta_{\rm M}$.
Integrating equation (\ref{Fr}) from $\theta_{\rm m}$ to $\theta_{\rm M}$,
we obtain $F(r,0)$ involving gravitational effects.
In Figure 5 we plot $F(r,0)$ for both the case when the bending is taken into
consideration and the case when it is not.
In comparison with the spherical case in Figure 2,
the deposited energy at distant regions in the disk case
is marginally substantial.
In the presence of bending, the peak of the energy deposition rate is
shifted to a little bit larger $r$  and the value of the rate at the peak
is about twice as larger as values obtained in the absence of bending.

Although the $\theta$ dependence of the gravitational effects may not necessarily
be small, unlike $F(r,\theta)$ in the absence of gravitation,
we assume it is small here. Using $F(r,0)$
with the bending effect, we calculate the energy deposition rate in the range
$\theta \le \pi/4$ and $r=2 r_g-10 r_g$.
Table 3 lists the values of the factor $f$ that shows the gravitational effects.
The $\theta$-dependence is neglected in our calculation
except for the case involving the redshift only.
Thus $f$ is normalized to unity when $F(r,\theta)$
(in the case involving the effect of the redshift only)
or $F(r,0)$ (in the other cases) is integrated over $r$ and $\theta$
in the absence of gravitation.
As is easily seen, the gravitational effects cancel one another out.
This is analogous to the neutrino sphere case in the previous section.
This result strongly supports that of Ruffert \& Janka (1999).
They treated the system of the accretion torus and a black hole unlike our
system of the disk and a black hole.
Using an approximation similar to ours,
they analytically calculated
the energy deposition rate due to the neutrino pair annihilation.
Their result is that the gravitational effects reduce the deposition rate by a factor
of $10-30\%$.
It agrees well with our result.

In order to circumvent the baryon contamination problem,
the energy fraction of baryonic matter in the fireball
must be less than about $10^{-5}$ (Shemi \& Piran 1990).
If we adopt the duration time of the neutrino radiation to be $t_{\rm
dur}=0.1$s and $T_{\rm eff}=10$MeV,
the highest mean mass densities $\bar{\rho}$ inside $\theta=0-\pi/3$ to resolve
the above problem are
$10^6$g/${\rm cm^3}$ for $r=2 r_g-5 r_g$,
$10^5$g/${\rm cm^3}$ for $r=5 r_g-10 r_g$ and
$10^4$g/${\rm cm^3}$ for $r=10 r_g-20 r_g$.
Since some fraction of energy really
escapes from the considered regions during the finite duration time,
the above restrictions may become more stringent.

\section{CONCLUSIONS}

\indent

In this article we have investigated semianalytically
the neutrino pair annihilation near
the neutrinosphere and around the thin accretion disk assuming that the
gravitational sources in both cases are described by the Schwarzschild metric.
The accretion disk has been assumed to be a blackbody and isothermal.
These assumptions enable us to treat these two cases based in an almost
unified fashion, which also clarifies the physical differences between these
two cases.
We have studied the general relativistic effects only near the rotation axis,
because that region is especially of interest to the source of GRBs
and estimating the effect of orbital bending
for large $\theta$ is difficult.

The general relativistic effects as a whole do not enhance the neutrino energy
deposition rate in either case.
The energy deposition rate is enhanced by
the effect of orbital bending toward the center.
However, the enhancement is cancelled out by the effects of the redshift and
capture by the gravitational attraction.
Consequently, numerical simulations of the neutrino energy
deposition rate in various models can correctly estimate
the order of the rate without considering the gravitational effects,
since it is supposed
that the thickness, shape, or temperature distribution of the disk or sphere
does not greatly affect the gravitational effects themselves.
Taking into account also the results of Ruffert \& Janka (1999), the
conclusions mentioned above are strongly suggested to be valid in the following
geometrical forms of the neutrino source: sphere, thin disk and torus.
We have also shown that the neutrinos emitted from the disk can
deposit energy at more distant regions than the neutrinos emitted
from the sphere.
The importance in this article resides in
the qualitative properties of the general relativistic effects.
The quantitative calculations in this paper are not so important,
and should be investigated on the basis of more sophisticated models and simulations.

\vspace{1.5cm}

We appreciate the helpful advice of M. Ruffert.
This work was partly supported by a Research Fellowship of the Japan Society for
the Promotion of Science.

\newpage

\begin{center}
{\bf \LARGE References}
\end{center}

\medskip

\begin{description}

\item
Berezinsky, V. S., \& Prilutsky, O. F. 1987, A\&A 175, 309
\item
Chandrasekhar, S. 1983, The Mathematical Theory of Black Holes
(New York: Oxford)
\item
Cooperstein, J., Van Den Horn, L. J., \& Baron, E. 1987, ApJ 321, L129
\item
Eichler, D., Livio, M., Piran, T., \& Schramm, D. N. 1989, Nature 340, 126
\item
Goodman, J., Dar, A., \& Nussinov, S. 1987, ApJ 314, L7
\item
Janka, H.-T., \& Ruffert, M. 1996, A\&A 307, L33
\item
Katz, J. I. 1997, ApJ 490, 633
\item
Landau, L. D., \& Lifshitz, E. M. 1979, Classical Theory of Fields
(London: Pergamon)
\item
MacFadyen, A., \& Woosley, S. E. 1999, ApJ 524, 262
\item
M\'esz\'aros, P., \& Rees, M. J. 1992a, ApJ 397, 570
\item
M\'esz\'aros, P., \& Rees, M. J. 1992b, MNRAS 257, 29p
\item
M\'esz\'aros, P., \& Rees, M. J. 1993, ApJ 405, 278
\item
Misner, C. W., Thorne, K. S., \& Wheeler, J. A. 1975,
Gravitation (New York: Freeman)
\item
Narayan, R., Paczy\'nski, B., \& Piran, T. 1992, ApJ 395, L83
\item
Paczy\'nski, B. 1990, ApJ 363, 218
\item
Paczy\'nski, B. 1998, ApJ 494, L45
\item
Paczy\'nski, B., \& Wiita, P. J. 1980, A\&A 88, 23
\item
Popham, R., Woosley, S. E., \& Fryer, C. 1999, ApJ 518, 356
\item
Rees, M. J., \& M\'esz\'aros, P. 1992, MNRAS 258, 41p
\item
Ruffert, M., \& Janka, H.-T. 1998, A\&A 338, 535
\item
Ruffert, M., \& Janka, H.-T. 1999, A\&A 344, 573
\item
Ruffert, M., Janka, H.-T., Takahashi, K., \& Sch\"afer, G. 1997, A\&A 319, 122
\item
Salmonson, J. D., \& Wilson, J. R. 1999, ApJ 517, 859
\item
Sari, R., Narayan, R., \& Piran, T. 1996 ApJ 473, 204
\item
Sari, R., \& Piran, T. 1995, ApJ 455, L143
\item
Shapiro, S. L., \& Teukolsky, S. A. 1983, Black Holes, White Dwarfs and Neutron Stars
(New York: Wiley)
\item
Shemi, A., \& Piran, T. 1990, ApJ 365, L55
\item
Woosley, S. E. 1993, ApJ 405, 273

\end{description}
\newpage

\begin{center}
{\bf \large Figure captions}
\end{center}

\figcaption{Maximum angle $\theta_{\rm M}$ at a distance $r$ from the
center formed by neutrinos emitted from the surface of the neutrinosphere.
The dashed line shows the orbit forming the maximum
angle in the absence of gravitation.}

\figcaption{The plot of $F(r)$ against $r$ for $R_{\nu}=2.5r_g$ when neutrinos are
emitted isotropically.
The solid (dashed) line shows $F(r)$ for the case in the presence (absence) of
gravitation. The origin is at $r=R_{\nu}$.}

\figcaption{Geometry of neutrino's path.
A neutrino is emitted from a point $S=(R,\pi/2,\varphi)$ on the disk and arrives at
$P=(r,\theta,0)$. The angular components of the vector joining $S$ and $P$ are
$(\theta_{\nu}, \varphi_{\nu})$.
$R$ (radius on the disk) is limited to the range from $R_{\rm in}$ to $R_{\rm out}$.}

\figcaption{Contour plot of $F(r,\theta)$ where neutrinos are
emitted from the disk. Here gravitation has been neglected.
(a) Plot in Cartesian coordinates $(r \sin{\theta},r \cos{\theta})$.
(b) Plot in polar coordinates, $(r, \theta)$.
The contours
are plotted at unit intervals except for the line of $F(r,\theta)=0.1$;
$\theta=0$ and $\theta=\pi/2$ correspond to points on the rotation
axis and on the disk, respectively.}

\figcaption{Diagram of $F(r,0)$ vs. $r$
where neutrinos are emitted from the disk.
The solid (dashed) line shows $F(r,0)$ for the case in which we
consider (neglect) the effect of gravitational bending.}

\newpage

\begin{center}
\begin{tabular}{ccccc}
\hline \hline
$R_{\nu}/r_{g}$ & \multicolumn{4}{c}{$f$} \\
&Redshift Only & Bending Only & Redshift and Bending & Whole \\ \hline
1.5 & 0.32 & 4.7 & 0.97 & 0.57 \\
2.5 & 0.60 & 1.6 & 0.87 & 0.73 \\
5 & 0.81 & 1.2 & 0.93 & 0.89  \\ \hline
\end{tabular}
\end{center}
{\footnotesize Table~1. The dimensionless factor $f$ represents
the general relativistic effects (see eq. [\ref{Gf}])
when neutrinos are emitted isotropically from the neutrinosphere;
$f$ is normalized to unity in the absence of gravitation.
The column headings "Redshift Only" and so on indicate
the incorporated effects of gravitation;
"Whole" over the last column means that we incorporate
all gravitational effects, redshift, bending, and trapping.
The energy deposition rate is enhanced by orbital bending
and reduced by the redshift and trapping.}

\begin{center}
\begin{tabular}{cccc}
\hline \hline
$\theta$ & $2 r_g-5 r_g$ & $5 r_g-10 r_g$ & $10 r_g-20 r_g$ \\ \hline
$0-\pi/4$ & 0.35 & 0.65 & 0.22 \\
$\pi/4-\pi/3$ & 0.32 & 0.77 & 0.17 \\ \hline
\end{tabular}
\end{center}
{\footnotesize Table~2. The dimensionless factor $G$.
It represents the fraction of the energy deposition rate for each region
(see eq. [\ref{Gf2}])
when neutrinos are emitted from the disk.
$G$ is normalized to unity for the region surrounded by
$r=2 r_g-10 r_g$ and $\theta=0-\pi/4$.
Here we neglect the effects of gravitation.}

\begin{center}
\begin{tabular}{ccccc}
\hline \hline
range & \multicolumn{4}{c}{$f$} \\
&redshift only & bending only & redshift and bending & whole \\ \hline
$2 r_g-5 r_g$ & 0.66 & 2.6 & 1.6 & 1.4 \\ 
$5 r_g-10 r_g$ & 0.31 & 2.5 & 0.78 & 0.75 \\ \hline
\end{tabular}
\end{center}
{\footnotesize Table~3. The dimensionless factor $f$ for $\theta=0-\pi/4$
(see equation [\ref{Gf2}]).
Neutrinos are emitted from the disk;
$f$ is normalized to unity when $F(r,\theta)$
(in the case involving the effect of the redshift only)
or $F(r,0)$ (in the other cases) is integrated over $r$ and $\theta$
in the absence of gravitation.}


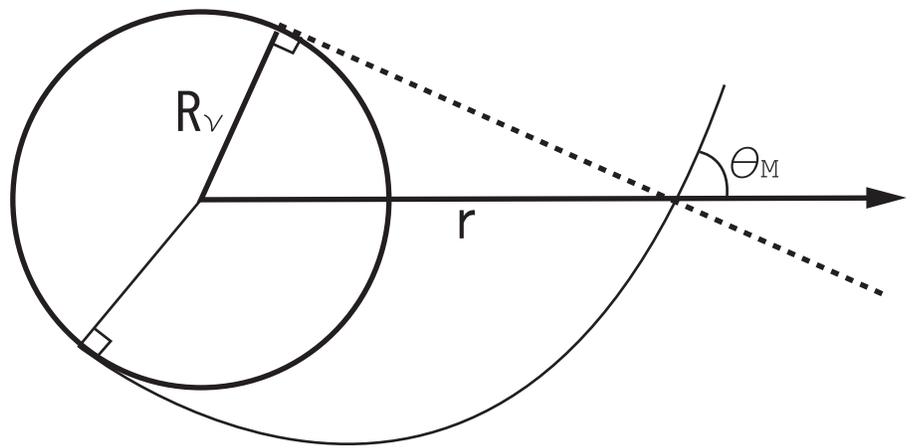
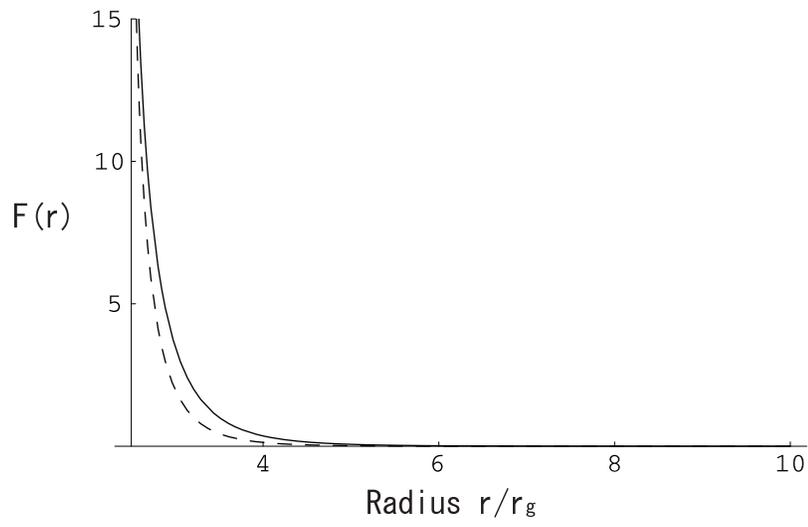
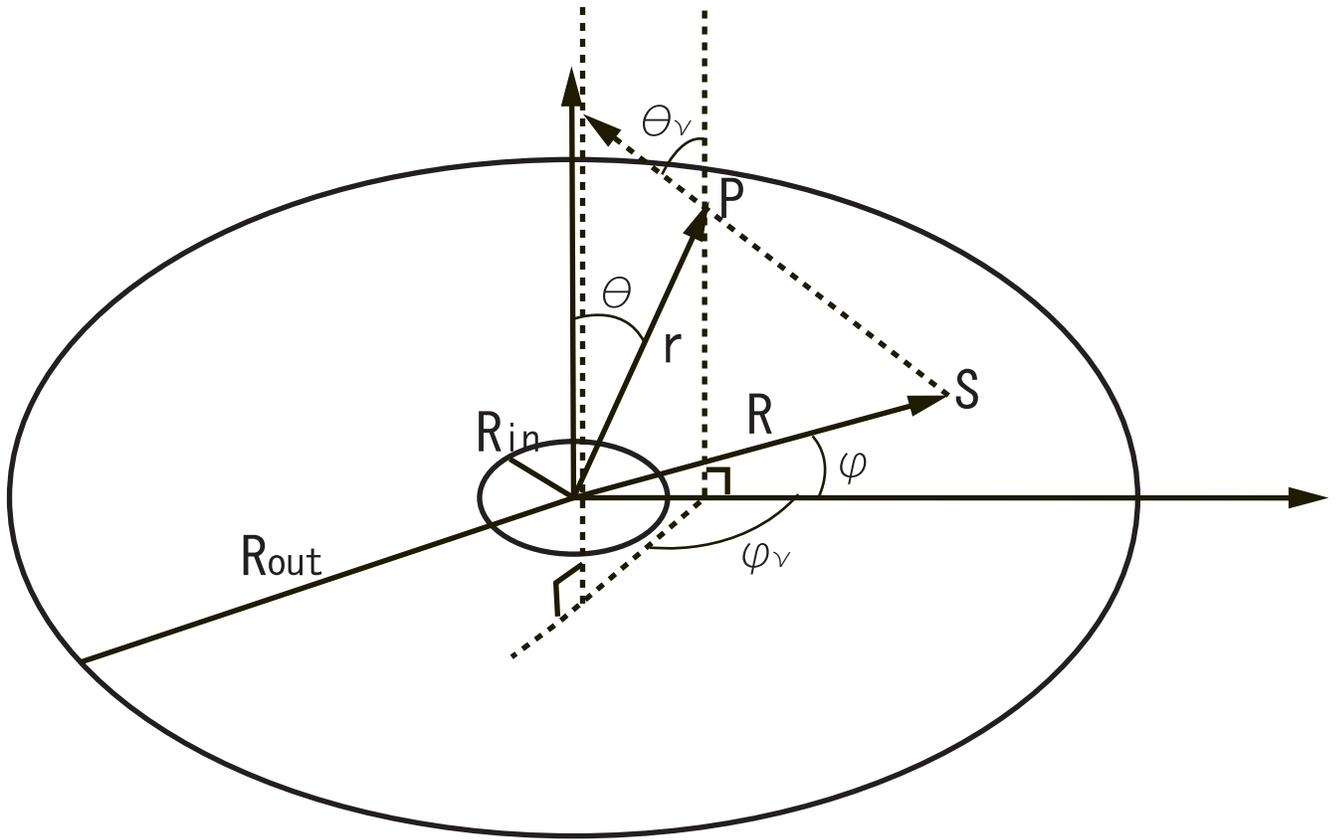
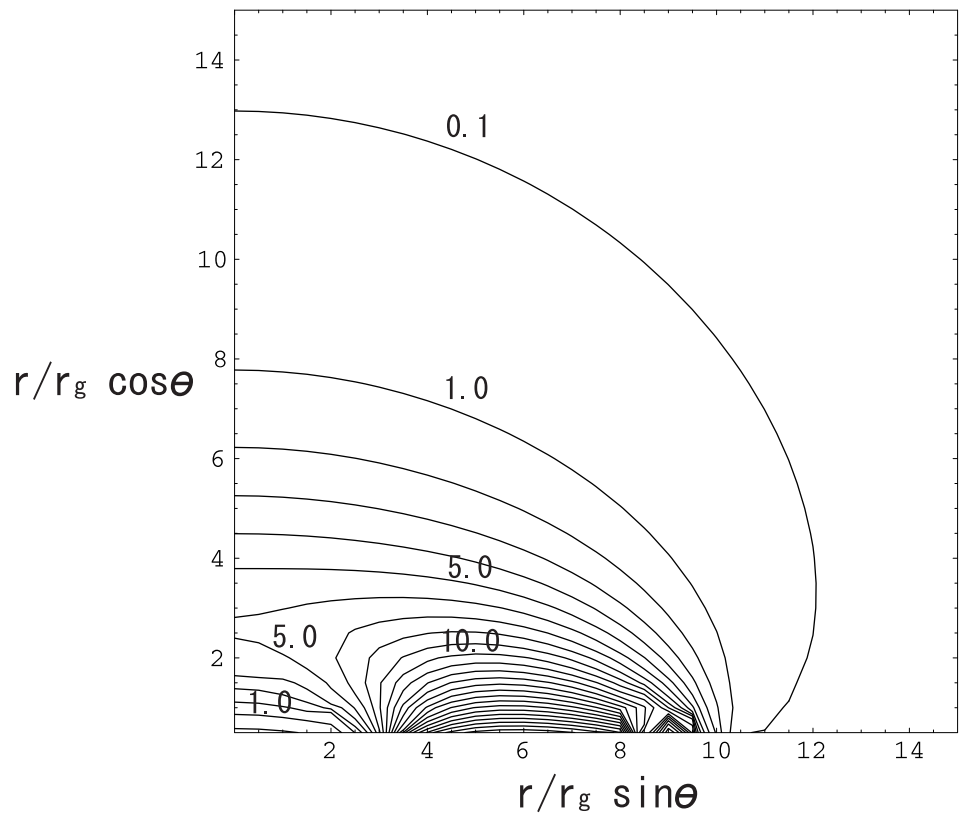
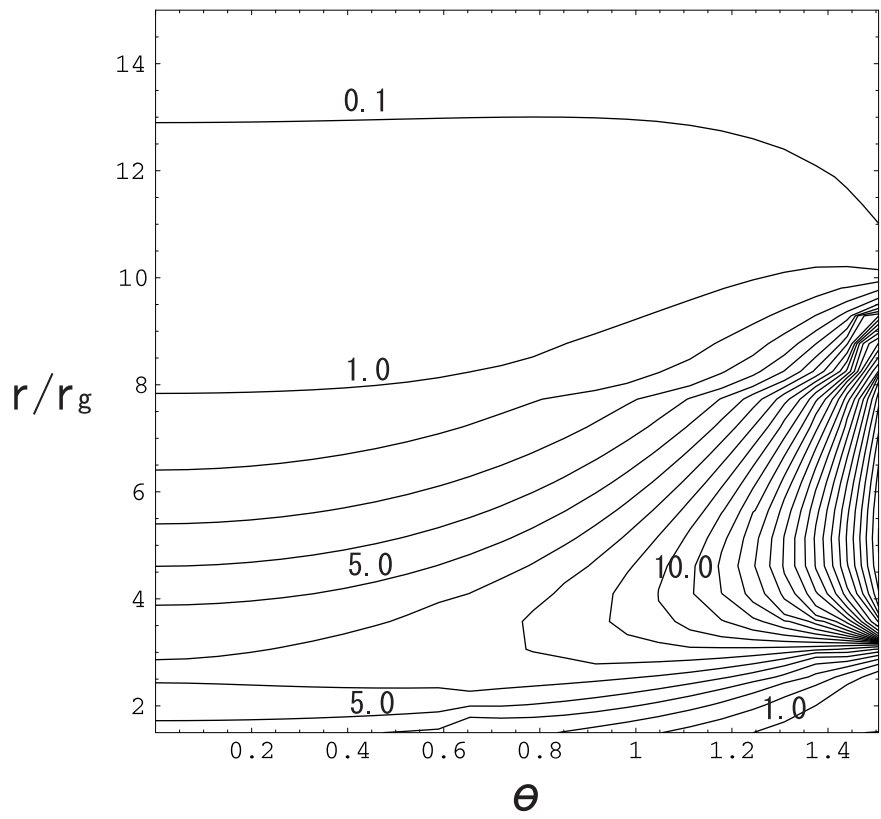
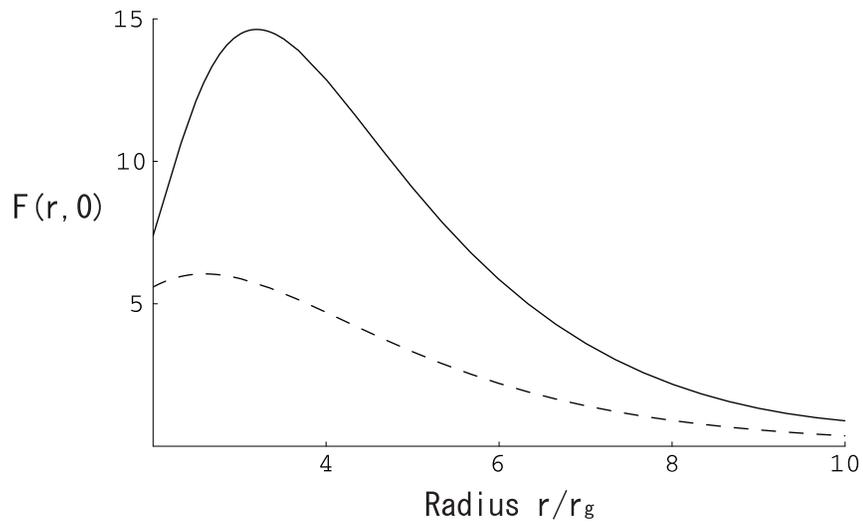
\end{document}